\documentclass[lettersize,journal,anonymous]{IEEEtran}
\usepackage{amsmath,amsfonts}
\usepackage{algorithmic}
\usepackage{algorithm}
\usepackage{array}
\usepackage[caption=false,font=normalsize,labelfont=sf,textfont=sf]{subfig}
\usepackage{textcomp}
\usepackage{stfloats}
\usepackage{url}
\usepackage{verbatim}
\usepackage{cite}
\usepackage{graphicx}
\usepackage{multirow}
\usepackage{xcolor}
\usepackage{hyperref}
\begin{document}

\title{An Empirical Study of Code Obfuscation Practices in the Google Play Store\textsuperscript{\dag{}}}

\author{Akila Niroshan,  
        Suranga Seneviratne,  
        Aruna Seneviratne
        
\thanks{\dag{} This is an extension of our previous work~\cite{akila2025stateofobfuscation} published in the ACM/SIGAPP Symposium on Applied Computing (SAC), 2025.}

\thanks{Akila Niroshan and Aruna Seneviratne are with the University of New South Wales (UNSW), Sydney, Australia (e-mail: a.pothpitiyage\_don@unsw.edu.au, a.seneviratne@unsw.edu.au).}
\thanks{Suranga Seneviratne is with the University of Sydney, Sydney, Australia (e-mail: suranga.seneviratne@sydney.edu.au).}

}

\IEEEpubid{\begin{minipage}{\textwidth}\centering
This work has been submitted to the IEEE for possible publication. Copyright~\copyright~ may be transferred without notice, after which this version may no longer be accessible.
\end{minipage}}
\IEEEpubidadjcol

\maketitle

\begin{abstract}
The Android ecosystem is vulnerable to issues such as app repackaging, counterfeiting, and piracy, threatening both developers and users. To mitigate these risks, developers often employ code obfuscation techniques. However, while effective in protecting legitimate applications, obfuscation also hinders security investigations as it is often exploited for malicious purposes. As such, it is important to understand code obfuscation practices in Android apps. In this paper, we analyze over 500,000 Android APKs from Google Play, spanning an eight-year period, to investigate the evolution and prevalence of code obfuscation techniques. First, we propose a set of classifiers to detect obfuscated code, tools, and techniques and then conduct a longitudinal analysis to identify trends. Our results show a 13\% increase in obfuscation from 2016 to 2023, with ProGuard and Allatori as the most commonly used tools. We also show that obfuscation is more prevalent in top-ranked apps and gaming genres such as Casino apps. To our knowledge, this is the first large-scale study of obfuscation adoption in the Google Play Store, providing insights for developers and security analysts.
\end{abstract}


\section{Introduction}
\label{Sec:Introduction}

\IEEEPARstart{A}{ndroid} plays a vital role in the smartphone market, holding more than 70\% market share~\cite{StatCounter_Global_Stats}. Within that, the Google Play Store serves as the primary app repository, offering over 1.7 million apps as of August 2024~\cite{a2024_google_number}. The Google Play Store is highly accessible, allowing developers to publish and monetize their apps with fewer barriers to entry.

Due to the relative ease of publishing apps, various malpractices are common in the Google Play Store~\cite{GARG2021102087}. For example, some malicious authors may repackage or counterfeit legitimate apps for nefarious purposes~\cite{zhou2012detecting, bhat2019survey}. Similarly, some malicious parties may steal intellectual property (IP) by reverse engineering apps~\cite{albakri2022survey}. These malpractices pose a significant threat to legitimate app developers and end users. As a countermeasure, app developers use code obfuscation to protect their apps and intellectual property (IP)~\cite{faruki2016android}. Equally, malware authors can also use obfuscation to evade anti-malware tools and to conceal functionality~\cite{elsersy2022rise, SIHAG2021100365}.

Although obfuscation provides security benefits, it also hinders reverse engineering, which is essential for static analysis by app investigators~\cite{molina2025light, beer2024tabbed, tan2023ptpdroid}. This poses challenges for malware analysts and app store administrators in enforcing security policies, as obfuscation can bypass anti-malware mechanisms and app store regulations~\cite{elsersy2022rise, SIHAG2021100365}. Additionally, obfuscation introduces performance degradations and limitations in Android research\cite{tan2023ptpdroid, gao2024comprehensive, pradeep2022not}. While some studies propose obfuscation-resilient research methods~\cite{liu2023enhancing, wang2022malwhiteout, wang2023uncovering}, others either overlook obfuscation or consider only basic techniques and tools\cite{li2024malcertain, liu2023no, chen2024attention}, possibly due to limited awareness of its prevalence. Given these challenges, it is important to examine the prevalence and trends of obfuscation in the Google Play Store. While some prior work has introduced methods for obfuscation detection~\cite{dong2018understanding, kuhnel2015fast, park2019framework, wang2017changed, wermke2018large}, no studies have fully examined the use of code obfuscation by app developers, the tools and techniques they employ, or the evolution of obfuscation adoption over time.

To this end, in this paper, we investigate code obfuscation adoption and practices in the Google Play Store from different aspects to benefit various stakeholders of the app market ecosystem. For developers, our work highlights the importance of obfuscation tools and industry standards to protect apps and IP. For researchers, we provide insights into industry trends. For malware analysts and app store administrators, our work underscores the necessity of robust security measures and regulations. Our research involves developing a set of classifiers to detect obfuscated code using various tools and techniques and analyzing trends over time by conducting a longitudinal study using data from two snapshots of the Google Play Store, taken five years apart. More specifically, we make the following contributions.

\begin{itemize}
\IEEEpubidadjcol
    \item We propose a bank of classifiers to detect whether an app is obfuscated and, if so, identify the obfuscation tools and techniques used. On our test set, we achieved 97\% accuracy in detecting obfuscation, 99\% accuracy in identifying the tool used for obfuscation, and 88\% accuracy in identifying the obfuscation technique. 
   
    \item Using these classifiers, we conduct a longitudinal study spanning eight years, from 2016 to 2023, to understand how code obfuscation practices have evolved in the Google Play Store. To the best of our knowledge, this is the first large-scale study of its kind to analyse over half a million Android applications.
    
    \item We show that overall code obfuscation in the Google Play Store increased by nearly 13\% from 2016 to 2023. We also find that ProGuard~\cite{proguard} and Allatori~\cite{allatori} are the two most commonly used tools by developers. Gaming apps tend to use obfuscation more than non-gaming apps, with Casino games showing the highest prevalence, at 80\% of apps obfuscated, and over 85\% using multiple techniques.

    \item We report a 28\% increase in obfuscation among top developers from 2018 to 2023 and an 11.7\% increase among developers with only one app. Over 90\% of the top 1,000 apps are obfuscated, with higher-ranked apps using multiple obfuscation techniques more frequently than lower-ranked ones. We further report that ProGuard is the most commonly used tool among lower-ranked apps.

\end{itemize}

The rest of the paper is organized as follows. In Section~\ref{sec:Obfuscation Intro}, we present background information such as common obfuscation tools and techniques. Section~\ref{sec:Methodology} details our obfuscation detection framework, and Section~\ref{sec:large-scale-analysis} describes our large-scale dataset. We present our findings on obfuscation trends in the Google Play Store in Section~\ref{sec:Results}. Section~\ref{Sec:Related Work} reviews related work, while Section~\ref{sec:discussion} discusses the implications and the limitations of our work and concludes the paper.

\section{Background}
\label{sec:Obfuscation Intro}

\subsection{Common Obfuscation Techniques}
\label{sec:obfustexchniques}

Obfuscation systematically converts the source code of the program into a form that is beyond human readability. This transformation maintains the application's functionality unchanged while altering the program's code. Several previous works studied and categorized obfuscation techniques~\cite{zhang2021android,conti2022obfuscation,guo2022survey,dong2018understanding,wermke2018large}. In the following, we describe some of the well-known obfuscation techniques used in Android apps.

\subsubsection{Identifier Renaming (IR)}

In \textit{Identifier Renaming}, identifiers in the code (e.g., class names, method names, and field names) are substituted with random characters or strings. This aims to make the code less readable by obfuscating readable information without changing the program logic. 

\subsubsection{Control Flow Modification (CF)}
The primary concept behind Control Flow Modification is to change the sequence of program execution, making it more difficult to understand and analyze. This technique is commonly used to protect software from reverse engineering and tampering. In ~\cite{guo2022survey} and ~\cite{zhang2021android}, various methods for achieving CF are discussed, and we outlined the popular techniques below.

\begin{itemize}
    \item \textbf{Control flow flattening} incorporates a construct that may include an infinite or finite loop with a termination condition. Within this construct, individual basic blocks are encapsulated as cases of a switch statement. While the original basic block is executed during runtime, the process of decompiling the switch case statement and restoring the initial code is challenging, due to the convoluted `\texttt{if}' and `\texttt{goto}' statements.
    
    \item \textbf{Call indirection} involves creating a new method to invoke the original method. Within the course of code execution, each method call is shadowed by this intermediary method, which, in turn, invokes the original method. It introduces complexity in the process of code restoration and impairs readability~\cite{zhang2021android,bacci2018a}. 
    
    \item \textbf{Reflection} is a technique in Java to alter the runtime behaviour of a program dynamically. It primarily leverages the \texttt{Java.lang.reflect.*} API, an integral component of the native Java library, to access and manipulate methods during program execution. 
    
    \item \textbf{Other} obfuscation methods involve techniques that sometimes overlap with the earlier methods, and as such, delineating boundaries between these techniques is challenging. For example, ~\cite{li2019obfusifier} introduces adding \texttt{`nop'} instructions and unconditional jumps. This is known as \textit{junk code insertion}. Moreover, developers can employ \textit{opaque predicates}, such as conditional statements or branches, to create a simulated branch~\cite{guo2022survey}, which constitutes a \textit{bogus control flow}. This practice generates two branches yielding the same outcome, with one branch containing the original code and the other comprising unreachable junk instructions. 
\end{itemize}
  
\subsubsection{String Encryption (SE)} Storing sensitive information or identifiable prompts in plain text strings within the source code may render the application vulnerable to third-party examination and reverse engineering. To avoid that, \textit{String Encryption} transforms human-readable strings within the code into human-unreadable character sequences.

\subsection{Commonly used Obfuscation Tools}
\label{sec:commontools}

Developers usually resort to tools to obfuscate code. Previous works~\cite{wang2017changed, dong2018understanding, mirzaei2019androdet, park2019framework} have reported multiple code obfuscation tools of various kinds, as we describe below.

\begin{itemize}

\item {\textbf{ProGuard}}~\cite{proguard} is an inbuilt and free obfuscator for Android Studio by GuardSquare. It can be easily activated by adding ProGuard rules in the \texttt{build.gradle} file. ProGuard can perform only Identifier Renaming and Code Optimization as specified in the user guide.

\item \textbf{Allatori}~\cite{allatori} is a commercial obfuscator by Smardec Inc. It is  offered as both paid and free educational versions with equal functionality. Integrating Allatori into an Android project is similar to ProGuard. However, it requires adding the Allatori \texttt{jar} file and configuration file in the \texttt{build.gradle}. It supports all three main obfuscation techniques discussed in Section~\ref{sec:obfustexchniques}.

\item \textbf{DashO}~\cite{dasho} is another commercial obfuscator by PreEmptive Inc. It is a paid tool, with the possibility of requesting a 7-day evaluation licence. Developers can use DashO UI to open source code files and enable necessary configurations. DashO UI will then add the required settings to the \texttt{build.gradle} file. It supports all three main obfuscation techniques described earlier.

\item \textbf{Obfuscapk}~\cite{obfuscapk} was initially developed as an open-source obfuscation tool for researchers to obfuscate Android applications. It implements all three techniques discussed in Section~\ref{sec:Obfuscation Intro}. As a validation dataset, we used the AndroOBFS dataset~\cite{androobfs}, which was obfuscated using ObfuscAPK. Further details on this dataset and its role in validating our method's performance will be discussed in Section~\ref{sec:building_dataset}.

\item \textbf{DexGuard}~\cite{dexguard} is an advanced paid version of ProGuard, also provided by GuardSquare. DexGuard implements all three techniques discussed earlier and also supports Runtime Application Self-Protection (RASP) for app hardening. We were unable to obtain a free or evaluation version. Therefore, we do not use it when building the training set later.

\end{itemize}

Table~\ref{tab:ob_tool_cap} summarizes the features of these obfuscation tools.

\begin{table}[h]
\caption{Summary of Android obfuscation tools}
\label{tab:ob_tool_cap}
\resizebox{\columnwidth}{!}{%
\begin{tabular}{lccc}
\hline
\textbf{Tool} & \textbf{\begin{tabular}[c]{@{}c@{}}Identifier\\ Renaming\end{tabular}} & \textbf{\begin{tabular}[c]{@{}c@{}}Control Flow\\ Modification\end{tabular}} & \textbf{\begin{tabular}[c]{@{}c@{}}String\\ Encryption\end{tabular}} \\ \hline
{ProGuard} & Yes & No & No \\
{DashO} & Yes & Yes & Yes \\
{Allatori} & Yes & Yes & Yes \\
{ObfuscAPK} & Yes & Yes & Yes \\
{DexGuard} & Yes & Yes & Yes \\ \hline
\end{tabular}%
}
\end{table}

\section{Obfuscation Detection Framework}
\label{sec:Methodology}

We developed a machine learning-based framework to detect whether an app is being obfuscated or not, followed by what tool it has used for obfuscation and, finally, what type of obfuscation(s) are present. Our framework consists of a bank of classifiers that use the same Android APK-level features.

\begin{figure*}[!h]
    \centering
    \subfloat[Training and testing]{%
        \includegraphics[width=\linewidth]{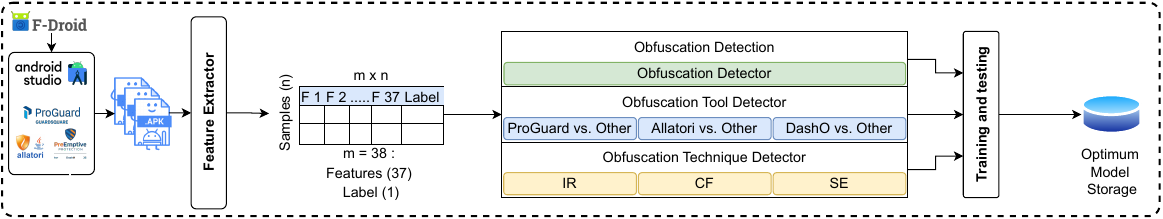}%
        \label{fig:overview_1a}
    }\\
    \subfloat[Large-scale analysis]{%
        \includegraphics[width=\linewidth]{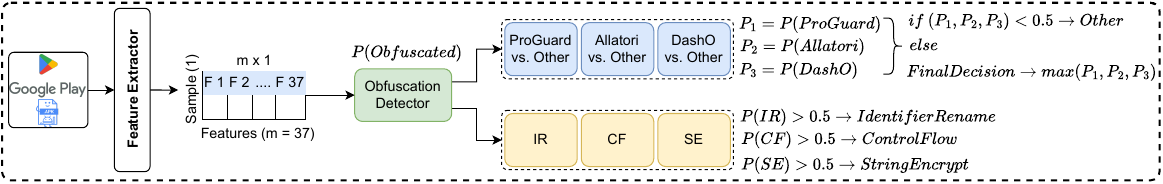}%
        \label{fig:overview_1b}
    }
    \caption{Overall experiment process: Training and testing and large-scale investigation.}
    \label{fig:overview}
\end{figure*}

\subsection{Classifier Banks}
\label{sec:classifier_bank}

Our framework consists of three classifier types as illustrated in Figure~\ref{fig:overview_1a} for  \textit{i) Obfuscation Detection, ii) Obfuscation Tool Detection, and iii) Obfuscation Technique Detection}.

Using our training and validation sets ({\bf cf.} Section~\ref{sec:building_dataset}), we tested several models, such as MLP, SVM, Random Forest, and Decision Trees. We selected the best model for each task and further tuned the hyper-parameters using grid search.

Our models and features are comparable to existing work in obfuscation detection~\cite{wermke2018large, wang2017changed, mirzaei2019androdet, conti2022obfuscation} and provide similar performance. Here, our methodological contribution is the comprehensive framework, which facilitates subsequent large-scale longitudinal analysis of obfuscation adoption.

\begin{itemize}
    \item \textbf{Obfuscation Detector}: Our Obfuscation Detector is a binary MLP classifier that makes a prediction of whether a given Android APK is obfuscated or not.

    \item \textbf{Obfuscation Tool Detector}: We use a bank of three binary Random Forest classifiers for obfuscation tool detection making decisions: \textit{1. ProGuard vs. Other}, \textit{2. Allatori vs. Other}, \textit{3. DashO vs. Other}.
    Each classifier assigns a probability to each tool: Proguard, DashO, and Allatori. Based on the highest probability, we decide which tool the APK uses. We chose to use a bank of classifiers rather than a single multi-class classifier to handle the other category more effectively, given the absence of training data for unknown obfuscation tools. This approach simplifies decision-making, as they do not need to distinguish between multiple classes. Also, it allows easy scalability by adding new classifiers as data from other tools becomes available.

    \item \textbf{Obfuscation Technique Detector}: Similarly, we train a bank of three binary Random Forest classifiers for Obfuscation Technique Detection. We focused on the most commonly used obfuscation techniques that were discussed in Section~\ref{sec:obfustexchniques} and in previous works~\cite{zhang2021android, conti2022obfuscation, guo2022survey, dong2018understanding, wermke2018large}; i) Identifier Renaming (IR), ii) Control Flow Modification (CF), and iii) String Encryption (SE). Given an APK, these classifiers predict whether it is obfuscated with each of these techniques. If the probability given by a classifier is higher than 0.5, we categorize the APK as obfuscated using the relevant technique.

\end{itemize}

\subsection{Feature Engineering}
\label{sec:features}

We use 37 features of three types to represent an Android APK. They are primarily related to obfuscation techniques we focus on, as summarised in Table~\ref{tab:feature_list}. These features were selected based on prior works~\cite{mirzaei2019androdet, conti2022obfuscation}. We use Androguard~\cite{desnos2018androguard} to extract Identifier names, Strings and Instructions from APKs' DEX files. We calculate the percentages of class names, method names, field names, and other strings based on their lengths, the presence of special characters, and the presence of numeric characters. Specifically, we count the occurrences of names with lengths of 1, 2, 3, 4, and greater than 4, as well as those containing special characters, numeric characters, or both. For example, Feature 1, described in Equation~\eqref{feature1}, represents the percentage of class names with a length of 1 relative to the total number of class names in a given APK. Equation~\eqref{feature5} calculates the percentage of class names containing special characters relative to the total number of class names. We performed similar calculations for other attributes, as detailed in Table 2. For instructions, we selected five specific instructions: \texttt{nop, goto, invoke, if, and move}. We then calculated the percentage of these selected instructions out of the total number of available instructions, as shown in Equation~\eqref{feature_ins}.

\begin{table}[]
\caption{Extracted feature list}
\label{tab:feature_list}
\resizebox{\columnwidth}{!}{%
\begin{tabular}{ccl}
\hline
\textbf{\begin{tabular}[c]{@{}c@{}}Obfuscation \\ Category\end{tabular}} & \textbf{Category} & \multicolumn{1}{c}{\textbf{Feature}} \\ \hline
\multirow{6}{*}{\begin{tabular}[c]{@{}c@{}}Identifier\\ Renaming\end{tabular}} & \multirow{2}{*}{\begin{tabular}[c]{@{}c@{}}Class\\ Names\end{tabular}} & \begin{tabular}[c]{@{}l@{}}Feature 1 - 5:\\ Percentage of class names of length 1, 2, 3, 4 and \textgreater{} 4\end{tabular} \\ \cline{3-3} 
 &  & \begin{tabular}[c]{@{}l@{}}Feature 6 - 8:\\ Percentage of class names containing special characters, \\numeric characters, or both\end{tabular} \\ \cline{2-3} 
 & \multirow{2}{*}{\begin{tabular}[c]{@{}c@{}}Method\\ Names\end{tabular}} & \begin{tabular}[c]{@{}l@{}}Feature 9 - 13:\\ Percentage of method names of length 1, 2, 3, 4 and \textgreater{} 4\end{tabular} \\ \cline{3-3} 
 &  & \begin{tabular}[c]{@{}l@{}}Feature 14 - 16:\\ Percentage of class names containing special characters, \\numeric characters, or both\end{tabular} \\ \cline{2-3} 
 & \multirow{2}{*}{\begin{tabular}[c]{@{}c@{}}Field\\ Names\end{tabular}} & \begin{tabular}[c]{@{}l@{}}Feature 17 - 21:\\ Percentage of field names of length 1, 2, 3, 4 and \textgreater{} 4\end{tabular} \\ \cline{3-3} 
 &  & \begin{tabular}[c]{@{}l@{}}Feature 22 - 24:\\ Percentage of class names containing special characters, \\numeric characters, or both\end{tabular} \\ \hline
\multirow{2}{*}{\begin{tabular}[c]{@{}c@{}}String\\ Encryption\end{tabular}} & \multirow{2}{*}{Strings} & \begin{tabular}[c]{@{}l@{}}Feature 25 - 29:\\ Percentage of other strings of length 1, 2, 3, 4 and \textgreater{} 4\end{tabular} \\ \cline{3-3} 
 &  & \begin{tabular}[c]{@{}l@{}}Feature 30 - 32:\\ Percentage of class names containing special characters, \\numeric characters, or both\end{tabular} \\ \hline
\begin{tabular}[c]{@{}c@{}}Control\\ Flow\end{tabular} & \multicolumn{1}{l}{Instructions} & \begin{tabular}[c]{@{}l@{}}Feature Ins 33 - 37:\\ Percentage of nop,  goto,  invoke,  if, and move instructions\end{tabular} \\ \hline
\end{tabular}%
}
\end{table}

\begin{equation}
    \label{feature1}
    \text{Feat. 1} = \frac{\text{No. of class names with length 1}}{\text {Total number of class names}}\times 100
\end{equation}

\begin{equation}
    \label{feature5}
    \text{Feat. 6} = \frac{\text{No. of class names consist of special chars}}{\text {Total number of class names}}\times 100
\end{equation}

\begin{equation}
    \label{feature_ins}
    \text{Feat. Ins 33} = \frac{\text{No. of \textbf{nop} Instructions}}{\text {Total number of Instructions}}\times 100
\end{equation}

\subsection{Building the Ground-truth Dataset}
\label{sec:building_dataset}

To build our training dataset, we downloaded app source codes from the F-droid repository~\cite{fdroid}. We imported each source code to Android Studio and disabled any obfuscation in the \texttt{build.gradle} to produce non-obfuscated samples.

To create obfuscated samples, we used the same projects imported from F-droid and employed ProGuard~\cite{proguard}, Allatori~\cite{allatori}, and DashO~\cite{dasho} as obfuscation tools. While Proguard is free, for Allatori, we used the educational version, which has the same features as the commercial version. For DashO, we used the 7-day evaluation licence provided to us by PreEmptive Inc., which again has the full features of the commercial version. To train the Obfuscation Technique detector, we created sets of APKs by applying each technique individually, thereby ensuring separate datasets for each technique.

Furthermore, we also use the AndroOBFS dataset~\cite{androobfs}, comprising malware APKs obfuscated with ObfuscAPK~\cite{obfuscapk}. Our aim is to select a subset of APKs from AndroOBFS to validate our classifiers and verify that they perform well with unseen obfuscation tools. Finally, we obtained a set of random APKs from the Google Play Store and manually labelled them for obfuscation, serving as an additional validation dataset. Since we don't know which tool was used to obfuscate these apps, we label them only as obfuscated or not.

Using manually created obfuscated and non-obfuscated APKs \textbf{(MC-APKs)}, AndroOBFS APKs, and 50 Google Play Store APKs \textbf{(GP)}, we generated several datasets to train and validate each classifier. The summary of these data subsets is discussed below.

\subsubsection{Datasets for Obfuscation Detector}

We curated four datasets using MC-APKs, AndroOBFS data, and 50 GP APKs:
\begin{itemize}
    \item {\bf D1}: Training and testing dataset (349 MC-APKs; 80\% for training and parameter tuning, and 20\% for testing). 
    \item {\bf D2:} Unseen evaluation dataset (135 MC-APKs).
    \item {\bf D3:} Random subset of AndroOBFS (270 APKs).
    \item {\bf D4:} Manually labelled 50 Google Play APKs.
\end{itemize}

Using F-Droid~\cite{fdroid} Android projects, we manually created 87 non-obfuscated APKs and 397 obfuscated APKs (MC-APKs). Due to the limited number of ground truth APKs, we divided these manually created APKs into two sets, D1 and D2, as detailed in Table~\ref{tab:obfuscation_detection_dataset}. We utilized D1 as our training and testing dataset, reserving D2 as an unseen validation dataset to assess the generalizability of our models. To further enhance the validation of generalizability, we incorporated the AndroOBFS dataset~\cite{androobfs}. We randomly selected D3 from the AndroOBFS dataset as our second validation set. As shown in Table~\ref{tab:obfuscation_detection_dataset}, all APKs in the AndroOBFS dataset are obfuscated; no non-obfuscated APKs were included. To strengthen the validation of our obfuscation detector, we also randomly selected 50 APKs from the Google Play Store and manually labelled them. We examined the identifier names of each APK to identify any anomalies or deviations in natural language. APKs were labelled as obfuscated if anomalies were observed in the identifier names; otherwise, they were labelled as non-obfuscated. Out of the 50 APKs, 33 were classified as obfuscated and 17 as non-obfuscated after manual labelling.

\begin{table}[!h]
    \centering
    \caption{Obfuscation Detector Dataset}
    \label{tab:obfuscation_detection_dataset}
    \begin{tabular}{p{2.0cm}ccc}
        \hline
        \textbf{Dataset} & \textbf{Obfuscated} & \textbf{N-Obfuscated} &\textbf{Total}\\
        \hline
        D1 & 274& 75& 349\\  
        D2 & 123& 12& 135\\ 
        D3 & 270& 0& 270\\ 
        D4 & 33& 17& 50\\
        \hline
    \end{tabular}
\end{table}

\subsubsection{Datasets for Obfuscation Tool Detector}

We use MC-APKs and AndroOBFS apps to create two datasets:
\begin{itemize}
    \item {\bf D5:} Training and testing dataset (312 MC-APKs; 80\% for training and parameter tuning, and 20\% for testing).
    \item {\bf D6:} Unseen evaluation dataset (50 MC-APKs + 30 AndroOBFS)
\end{itemize}

We excluded all non-obfuscated APKs from the manually created APK dataset (from MC-APKs) because they could not be categorized under any obfuscation tools. Additionally, we removed APKs that were obfuscated using a combination of two tools, as it was challenging to classify such APKs under a single tool. Consequently, we had 362 obfuscated APKs available for tool detection. Similar to the previous scenario, we divided this dataset into D5 and D6, using D5 for training and testing the tool detection model, and keeping D6 as the unseen validation set. To ensure that our model can accurately classify the ``Other'' category, we randomly selected 30 APKs from the AndroOBFS dataset and combined them with D6, as detailed in Table~\ref{tab:tool_detection_dataset}. We used D6 as our unseen validation dataset.

\begin{table}[!h] 
    \caption{Obfuscation Tool Detector Dataset}
    \label{tab:tool_detection_dataset}
    \begin{tabular}{p{1.5cm}ccccc} \hline
        Dataset & ProGuard & DashO & Allatori & Obfuscapk & Total\\ \hline
        D5 & 68 & 162 & 82 & 0 & 312\\
        D6 & 15& 20 & 15& 30& 80\\
        \hline
    \end{tabular}
\end{table}

\subsubsection{Datasets for Obfuscation Technique Detector}

We used 376 MC-APKs and AndroOBFS apps to create three datasets:
\begin{itemize}
    \item {\bf D7:} Training and testing dataset (324 MC-APKs; 80\% for training and parameter tuning, and 20\% for testing).
    \item {\bf D8:} Unseen evaluation dataset (52 MC-APKs)
    \item {\bf D9:} Random subset of AndroOBFS (90 each for IR, CF, and SE).
\end{itemize}
In line with the tool detection process, we removed non-obfuscated APKs and irregularly combined APKs from the original dataset (from MC-APKs) to create a set of 376 manually obfuscated APKs for technique detection. As with the previous cases, we split this dataset into two subsets: D7 for training and testing, and D8 for unseen evaluation. The number of APKs in each dataset is detailed in Table~\ref{tab:technique_detection_dataset}. Additionally, since the AndroOBFS dataset provides APKs labelled with obfuscation techniques, we randomly selected an extra set of APKs (D9) from it to further validate the generalizability of our method.

\begin{table}[!h] 
    \caption{Obfuscation Technique Detector ataset}
    \label{tab:technique_detection_dataset}
    \begin{tabular}{p{0.8cm}cccccccc} \hline
        Dataset & IR & CF & SE & IR\&CF & IR\&SE & CF\&SE & All & \textbf{Total}\\
        \hline
        D7 & 45 & 36 & 36 & 23 & 23 & 23 & 138 & 324\\
        D8 & 15 & 7 & 7 & 2 & 2 & 4 & 15 & 52\\
        D9 & 90 & 90 & 90 & - & - & - & - & 270\\
        \hline
    \end{tabular}
\end{table}

Note that there are different numbers of APKs in the MC-APKs dataset for the three detectors. This is because we excluded all non-obfuscated APKs and APKs created using multiple tools (i.e., ProGuard and Allatori combined) from the MC-APKs when creating the tool detector and technique datasets. Additionally, to balance the unseen evaluation set, we used only 30 APKs from AndroOBFS in D6.

\subsection{Performance of the Classifiers}
\label{sec:training_and_validation}

\subsubsection{{Obfuscation Detector}}

We show the performance of the obfuscation detector on different datasets in Table~\ref{tab:binary_cls_results}. Our results are comparable to those of OBFUSCAN~\cite{wermke2018large}, which reported similar findings. However, OBFUSCAN targets only ProGuard-obfuscated APKs, whereas our tool can handle APKs obfuscated by various obfuscators. To assess how well our obfuscation detector works with unseen data (i.e., not from a split of the training and test set), we evaluated it on D2, D3 and D4 as well. While the performance dropped somewhat, detector accuracy was still in the range of 87\%--92\%, suggesting its suitability for the large-scale analysis.

\begin{table}[h]
\caption{Obfuscation detection - Results}
\label{tab:binary_cls_results}
\resizebox{\columnwidth}{!}{%
\begin{tabular}{ccccc}
\hline
\textbf{Dataset} & \textbf{Accuracy} & \textbf{Precision} & \textbf{Recall} & \textbf{F1 Score} \\ \hline
D1 (Test Set) & 0.97 & 0.96 & 1.00 & 0.98 \\
D2 & 0.87 & 1.00 & 0.85 & 0.92 \\
D3 & 0.88 & 1.00 & 0.88 & 0.94 \\
D4 & 0.92 & 1.00 & 0.88 & 0.94 \\ \hline
Avg. & 0.91 & 0.99 & 0.90 & 0.95 \\ \hline
\end{tabular}%
}
\end{table}

\subsubsection{{Obfuscation Tool Detector Bank}}

Each classifier in our bank determines if an APK is obfuscated using a specific tool (ProGuard, Allatori, DashO) or another tool (Other). Here, it is crucial to assess that each classifier is not only classifying its own target tool usage correctly but also any other tool usage must be classified as other. For instance, if an APK is obfuscated with \emph{DashO}, \textit{ProGuard vs. Other} and \textit{Allatori vs. Other} should classify it as \emph{Other}, while \textit{DashO vs. Other} should detect it as \emph{DashO}. For APKs from AndroOBFS, all three classifiers should classify them as \emph{Other}. Therefore, we use macro versions of Precision, Recall, and F1 to evaluate performance. We summarize the results in Table~\ref{tab:tool_results}.

On the test set (D5), our classifier bank achieves an average of 99\% which is comparable to previous work~\cite{wang2017changed}. However, we highlight that~\cite{wang2017changed} operates in a closed-set setting and, as such, does not have the means to categorize unknown obfuscators accurately. To further validate how well the classifiers perform in detecting unknown tools, we evaluated them on D6. We achieved an average accuracy of 88\% showing that the obfuscation tool detector indeed works well with unknown tools and classifies them as others with high accuracy.

\begin{table}[h]
\caption{Obfuscation tool detection - Results}
\label{tab:tool_results}
\resizebox{\columnwidth}{!}{%
\begin{tabular}{clcccc}
\hline
\multirow{2}{*}{\textbf{Dataset}} & \multicolumn{1}{c}{\multirow{2}{*}{\textbf{Classifier}}} & \multicolumn{3}{c}{\textbf{Macro}} & \multirow{2}{*}{\textbf{Acc.}} \\ \cline{3-5}
 & \multicolumn{1}{c}{} & \textbf{P} & \textbf{R} & \textbf{F1} &  \\ \hline
\multirow{3}{*}{\begin{tabular}[c]{@{}c@{}}D5 \\ (Test Set)\end{tabular}} & ProGuard vs. Other & 1.00 & 1.00 & 1.00 & 1.00 \\
 & Allatori vs. Other & 1.00 & 1.00 & 1.00 & 1.00 \\
 & DashO vs. Other & 0.98 & 0.97 & 0.97 & 0.97 \\ \hline
\multirow{3}{*}{D6} & ProGuard vs. Other & 0.82 & 0.91 & 0.86 & 0.90 \\
 & Allatori vs. Other & 0.89 & 0.91 & 0.90 & 0.94 \\
 & DashO vs. Other & 0.84 & 0.93 & 0.87 & 0.89 \\ \hline
\end{tabular}%
}
\end{table}

\subsubsection{{Obfuscation Technique Detector Bank}}

Similarly, we show the performance of the obfuscation technique detector bank in Table~\ref{tab:technique_results}. On the test set (D7), we achieved an average accuracy of 88\% which is comparable to prior works~\cite{kuhnel2015fast, conti2022obfuscation, mirzaei2019androdet}. Our technique outperforms the method described in~\cite{mirzaei2019androdet}. While~\cite{kuhnel2015fast} focuses solely on identifier renaming, our method addresses a broader range of popular obfuscation techniques. Additionally, although the results of AndrODet*~\cite{conti2022obfuscation} for the training and test datasets are comparable to those of our method, they retrained their model before evaluating it on unseen data. This approach converts the unseen data into seen data, which undermines the generalizability of their method. Such generalizability is crucial for conducting a comprehensive large-scale study.

Further, to examine the generalizability of our method, we conducted two validation studies using D8 and D9, which are unseen and differently distributed from the original training/test dataset. Still, the classifier bank achieves average accuracies of 88\% and 80\%.

\begin{table}[h]
\caption{Obfuscation technique detection - Results}
\label{tab:technique_results}
\resizebox{\columnwidth}{!}{%
\begin{tabular}{cccccc}
\hline
\textbf{Dataset} & \textbf{Classifier} & \textbf{Precision} & \textbf{Recall} & \textbf{F1} & \textbf{Accuracy} \\ \hline
\multirow{3}{*}{D7 (Test Set)} & (IR) & 0.98 & 0.89 & 0.93 & 0.91 \\
 & (CF) & 0.87 & 0.91 & 0.89 & 0.85 \\
 & (SE) & 0.92 & 0.92 & 0.92 & 0.88 \\ \hline
\multirow{3}{*}{D8} & (IR) & 1.00 & 0.88 & 0.94 & 0.92 \\
 & (CF) & 0.82 & 0.96 & 0.89 & 0.86 \\
 & (SE) & 0.83 & 0.93 & 0.88 & 0.86 \\ \hline
\multirow{3}{*}{D9} & (IR) & 1.00 & 0.84 & 0.92 & 0.84 \\
 & (CF) & 1.00 & 0.78 & 0.88 & 0.78 \\
 & (SE) & 1.00 & 0.79 & 0.88 & 0.79 \\ \hline
\end{tabular}%
}
\end{table}

In summary, we presented an obfuscation detection framework comprising three machine learning models developed to detect obfuscation, obfuscation tools, and techniques. We evaluated the capabilities of these models to perform large-scale analysis using various ground-truth datasets, which are employed in the large-scale analysis as illustrated in Figure~\ref{fig:overview_1b}. In addition, we have publicly released the source codes of our classifiers along with its best models and ground-truth datasets.\footnote{\href{https://github.com/NSS-USYD/Obfuscation-Large_Analysis}{https://github.com/NSS-USYD/Obfuscation-Large\_Analysis}}

\section{Large Scale Analysis}
\label{sec:large-scale-analysis}

\subsection{Dataset}
\label{subsec:dataset}

\begin{table*}[h]
\tiny
\caption{Number of APKs per year}
\label{tab:large-scale_dataset}
\resizebox{\linewidth}{!}{%
\begin{tabular}{c|c|c|c|c|c|c|c|c|c}
\hline
\textbf{Year} & 2016 & 2017 & 2018 & 2019 & 2020 & 2021 & 2022 & 2023 & \textbf{Total} \\ \hline
\textbf{\begin{tabular}[c]{@{}c@{}} Available APKs\end{tabular}} & 174,136 & 501,865 & 157,613 & 7,205 & 21,014 & 60,705 & 240,775 & 65,697 & 1,229,010 \\ \hline
\textbf{Analysed APKs} & 74,817 & 159,639 & 80,112 & 7,201 & 20,982 & 59,539 & 81,134 & 65,543 & 548,967 \\ \hline
\end{tabular}%
}
\end{table*}

Our large-scale analysis data is based on two large snapshots of the Google Play Store collected around 2018 and 2023. The 2018 dataset that was collected as a part of our previous work~\cite{karunanayake2020multi,rajasegaran2019multi} contains metadata of over 1.2 million apps that were collected between January and March 2018 and 1,023,521 APK files. The 2023 dataset contains metadata of over one million apps and was collected between January and November and 395,396 APK files.

Both datasets were collected in the same way using a Python-based crawler. First, the crawler discovered available apps on the Google Play Store. Then, it collected app metadata (e.g., app ID, app genre, developer name, number of downloads, rating details) and APK executables for free apps. There are several reasons behind the difference between the number of apps for which we crawled metadata and the number of apps for which we downloaded the APKs. First, the APK crawler is significantly slower than the metadata crawler. Second, we do not download the APKs of paid apps. Third, some apps do not support the Android device we simulated to download APKs.

One of the fields in app metadata is the ``\texttt{last update date}''. We use this field to categorise apps by year as summarised in Table~\ref{tab:large-scale_dataset}. As can be seen, the centre years of our two crawls, i.e., 2017 and 2022, have the highest number of apps. We have a notably smaller number of apps for 2019 and 2020 because they were only collected in 2022, and only a limited number of apps have the last update date in 2019 and 2020. These apps can bias our analysis as these represent apps that have been most likely abandoned by app developers. As a result, we do not consider 2019 and 2020 in our extended analysis. For each year, we analyse a random sample of apps as listed in Table~\ref{tab:large-scale_dataset}. The reason for not analysing all apps in all the years is the time, as APK decompilation takes time.

\subsection{Process of APK Analysis} For each APK we analyse, we use Androguard~\cite{desnos2018androguard} and our pre-processing scripts to create the feature vector described in Section~\ref{sec:features}. Next, we make a prediction using our \textit{Obfuscation Detector}. If the app is predicted as not obfuscated, we record this and stop further analysis for that app. If the APK is obfuscated, we use the same feature vector with the \textit{Obfuscation Tool Detector Bank} and \textit{Obfuscation Technique Detector Bank} to identify the tool and technique(s) used. In the Tool Detector step, if all three classifiers give a probability of less than $0.5$, we categorize the APK as using an \textit{Other} tool. Otherwise, we use the highest probability to determine the tool. In the Technique Detector, the classifier identifies the obfuscation technique \textit{(IR, CF, SE)} if its probability exceeds $0.5$. This overall process is illustrated in Figure~\ref{fig:overview_1b}.
\section{Results}
\label{sec:Results}

In this section, we present various analysis results that demonstrate the adoption of code obfuscation in Google Play.

\subsection{Overall Obfuscation Trends} 
\label{sec:obstrend}

\subsubsection{Presence of obfuscation} 
\begin{figure}[h!]
\centering
    \includegraphics[width=\linewidth]{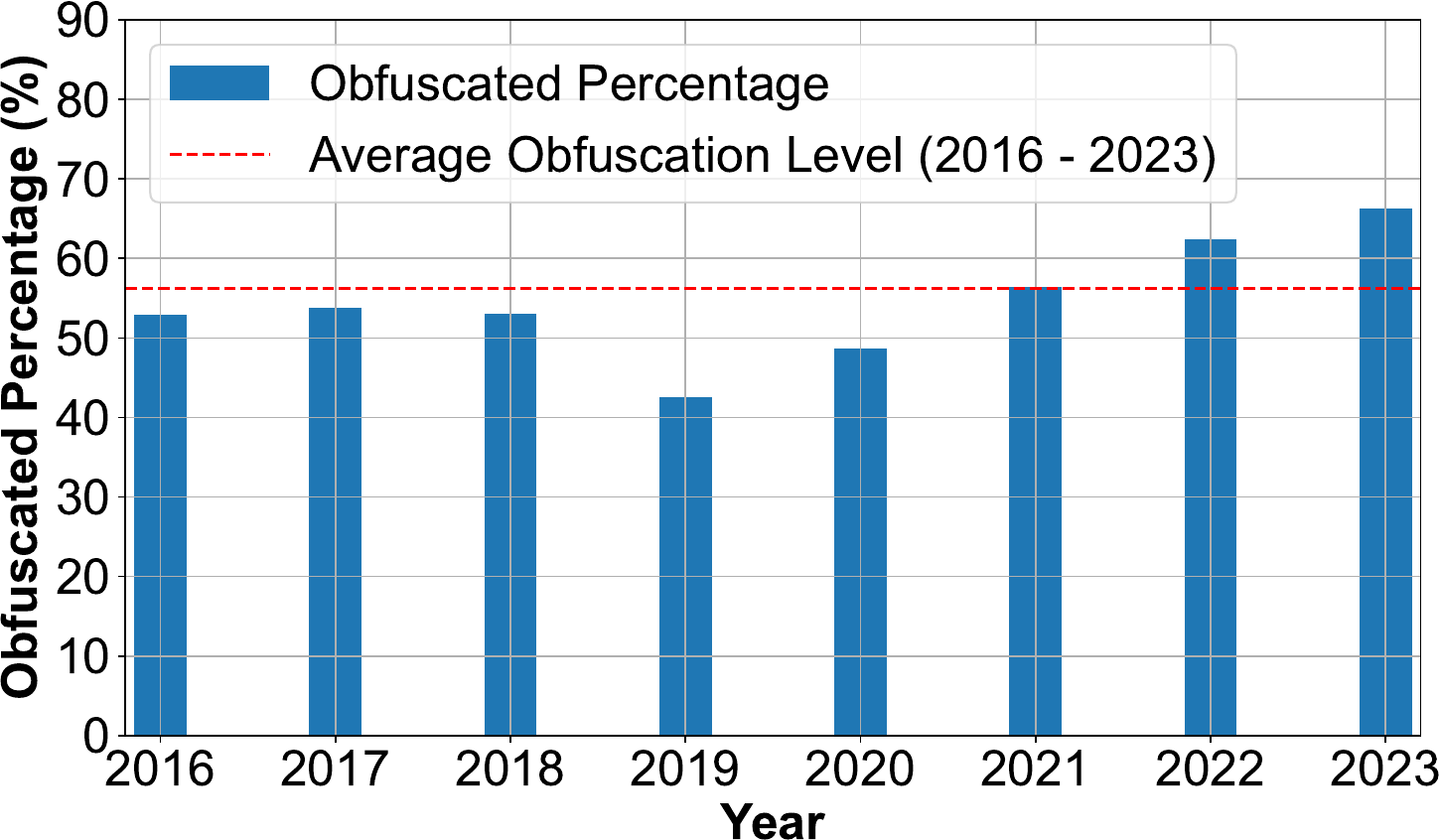}
    \caption{Percentage of obfuscated apps by year} \vspace{-4mm}
    \label{fig:obfuscated_percentage}
\end{figure}

Out of the 548,967 Google Play Store APKs analyzed, we identified 308,782 obfuscated apps, representing approximately 56.25\% of the total. In Figure~\ref{fig:obfuscated_percentage}, we show the year-wise percentage of obfuscated apps for 2016-2023. There is an overall obfuscation increase of 13\% between 2016 and 2023, and as can be seen, the percentage of obfuscated apps has been increasing in the last few years, barring 2019 and 2020. As explained in Section~\ref{subsec:dataset}, 2019 and 2020 contain apps that are more likely to be abandoned by developers, and as such, they may not use advanced development practices.

From 2016 to 2018, the obfuscation levels were relatively stable at around 50-55\%, while from 2021 to 2023, there was a marked rise, reaching approximately 66\% in 2023. This indicates a growing focus on app protection measures among developers, likely driven by heightened security and IP concerns and the availability of advanced obfuscation tools.

\subsubsection{Obfuscation tools} 
\begin{figure}[h]
\centering
    \includegraphics[width=\linewidth]{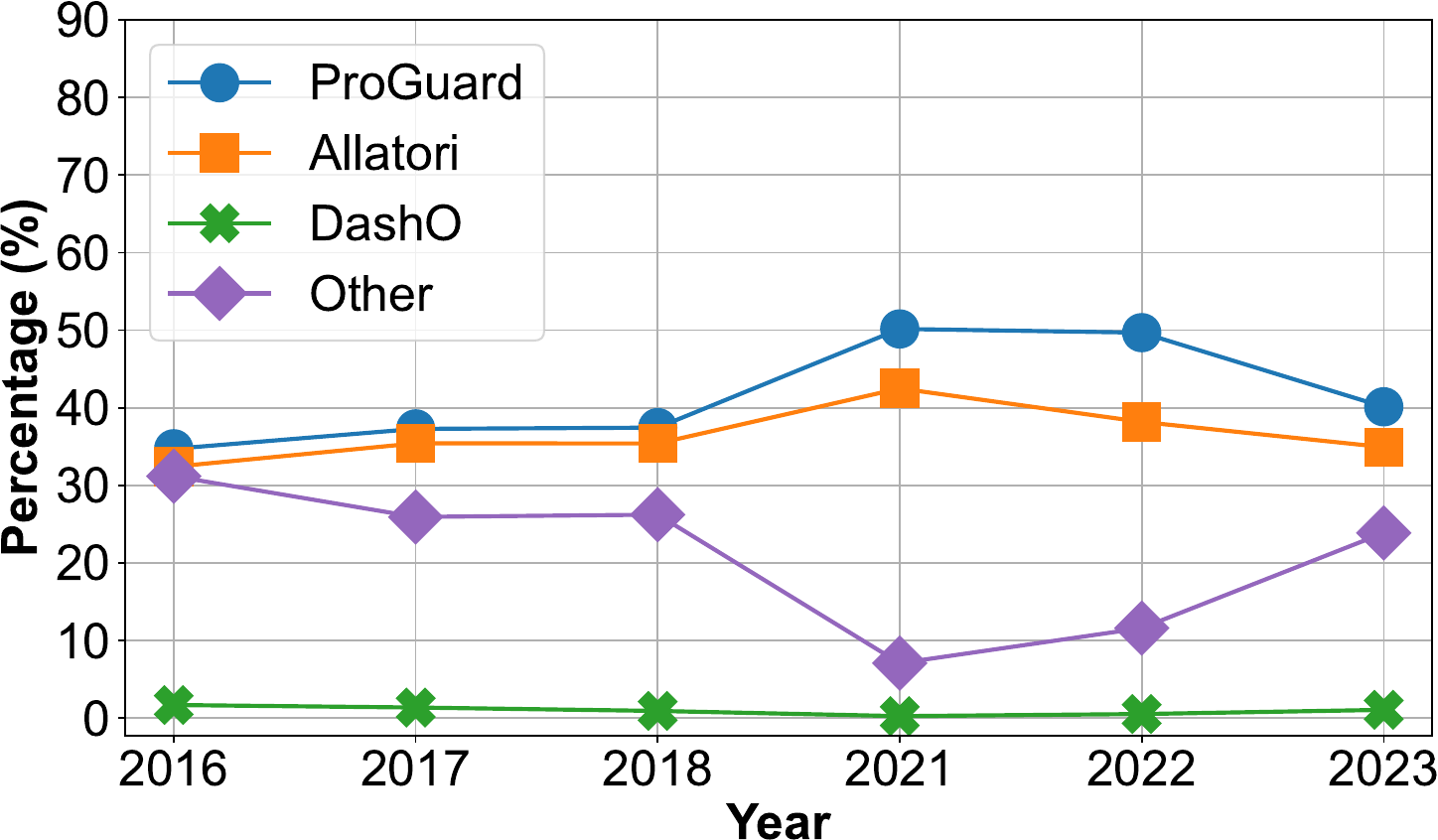} 
    \caption{Yearly obfuscation tool usage}
    \label{fig:ofbuscated_tool}
\end{figure}

Among the obfuscated APKs, our tool detector identified that 40.92\% of the apps use Proguard, 36.64\% use Allatori, 1.01\% use DashO, and 21.43\% use other (i.e., unknown) tools. We show the yearly trends in Figure~\ref{fig:ofbuscated_tool}. Note that we omit results in 2019 and 2020 ({\bf cf.} Section~\ref{subsec:dataset}).

ProGuard and Allatori are the most consistently used obfuscation tools, with ProGuard showing a slight overall increase in popularity and Allatori demonstrating variability. This inclination could be attributed to ProGuard being the default obfuscator integrated into Android Studio, a widely used development environment for Android applications. Notably, ProGuard usage increased by 13\% from 2018 to 2021, likely due to the introduction of R8 in April 2019~\cite{release_note_android}, which further simplified ProGuard integration with Android apps.

DashO consistently remains low in usage, likely due to its high cost. The use of other obfuscation tools decreased until 2018 but has shown a resurgence from 2021 to 2023. This suggests that developers might be using other or custom tools, or our detector might be predicting some apps obfuscated with Proguard or Allatori as `other.' To investigate, we manually checked a sample of apps from the `other' category and confirmed they are indeed obfuscated. However, we could not determine which obfuscation tools the developers used. We discuss this potential limitation further in Section~\ref{sec:limitations}.

\subsubsection{Obfuscation techniques} 
\begin{figure}[h]
\centering
    \includegraphics[width=\linewidth]{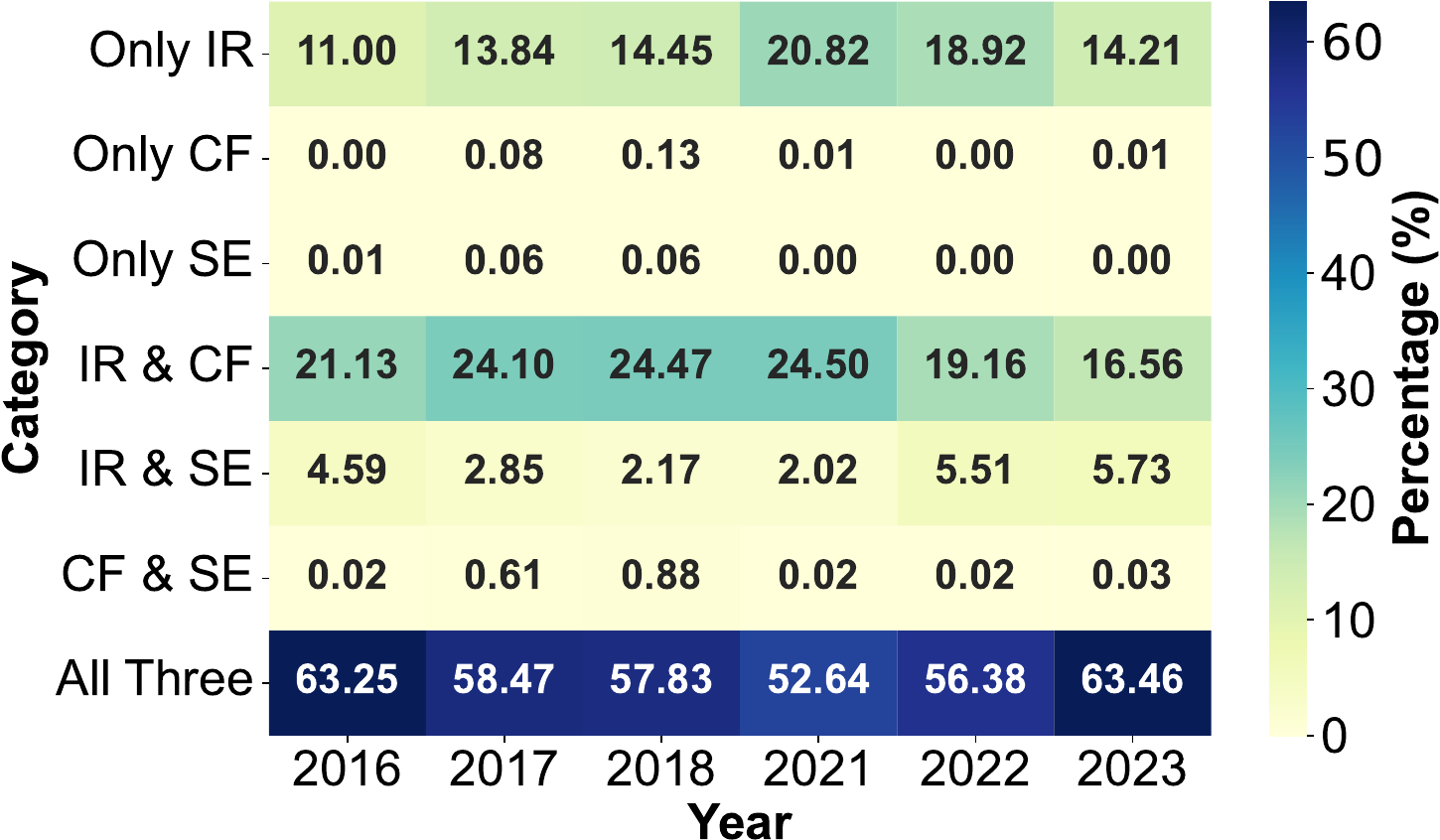} 
    \caption{Yearly obfuscation technique usage}
    \label{fig:obfuscated_tech}
\end{figure}

We show the year-wise breakdown of obfuscation technique usage in Figure~\ref{fig:obfuscated_tech}. Among the various obfuscation techniques, Identifier Renaming emerged as the most prevalent, with 99.62\% of obfuscated apps using it alone or in combination with other methods (Categories of Only IR, IR and CF, IR and SE, or All three). Furthermore, 81.04\% of obfuscated apps used Control Flow Modification, and 62.76\% used String Encryption. The pervasive use of Identifier Renaming (IR) can be attributed to the fact that all obfuscation tools support it ({\bf cf.} Table~\ref{tab:ob_tool_cap}). Similarly, lower adoption of Control Flow Modification and String Encryption can be attributed to Proguard not supporting it. 

Next, we investigate the adoption of obfuscation on Google Play Store from various perspectives. Same as earlier, due to the smaller dataset size and possible bias ({\bf cf.} Section~\ref{subsec:dataset}), we exclude the APKs from 2019 and 2020 from this analyses.

\subsection{App Genre}
\label{sec:app_genre}

First, we investigate whether the obfuscation practices vary according to the App genre. Initially, we analysed all the APKs together before separating them into two snapshots.

\begin{figure*}[h]
    \centering
    \includegraphics[width=\linewidth]{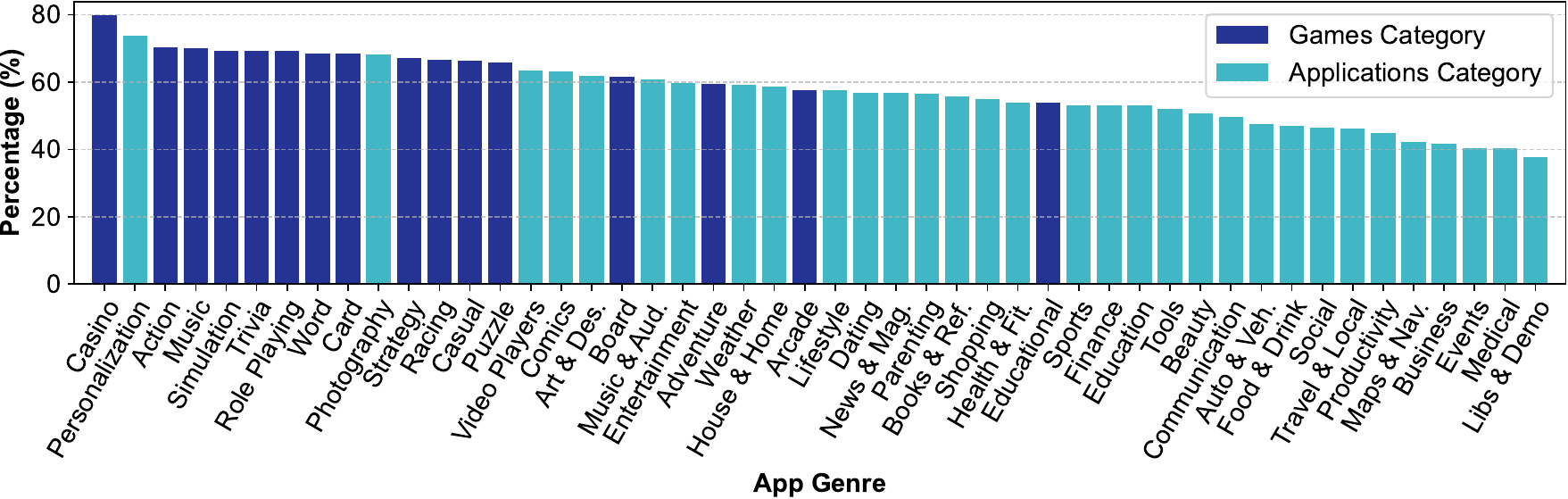}
    \caption{Obfuscated app percentage by genre (overall)}
    \label{fig:app_genre_overall}
\end{figure*}

Figure~\ref{fig:app_genre_overall} shows the genre-wise obfuscated app percentage. We note that 19 genres have more than 60\% of the apps obfuscated, and almost all the genres have more than 40\% obfuscation percentage. \textit{Casino} genre has the highest obfuscation percentage rate at 80\%, and overall, game genres tend to be more obfuscated than the other genres. The higher obfuscation usage in casino apps is logical due to their nature. These apps often simulate or involve gambling activities and handle monetary transactions and sensitive data related to in-game purchases, making them attractive targets for reverse engineering and hacking. This necessitates robust security measures to prevent fraud and protect user data.

\begin{figure}[h]
    \centering
    \includegraphics[width=\linewidth]{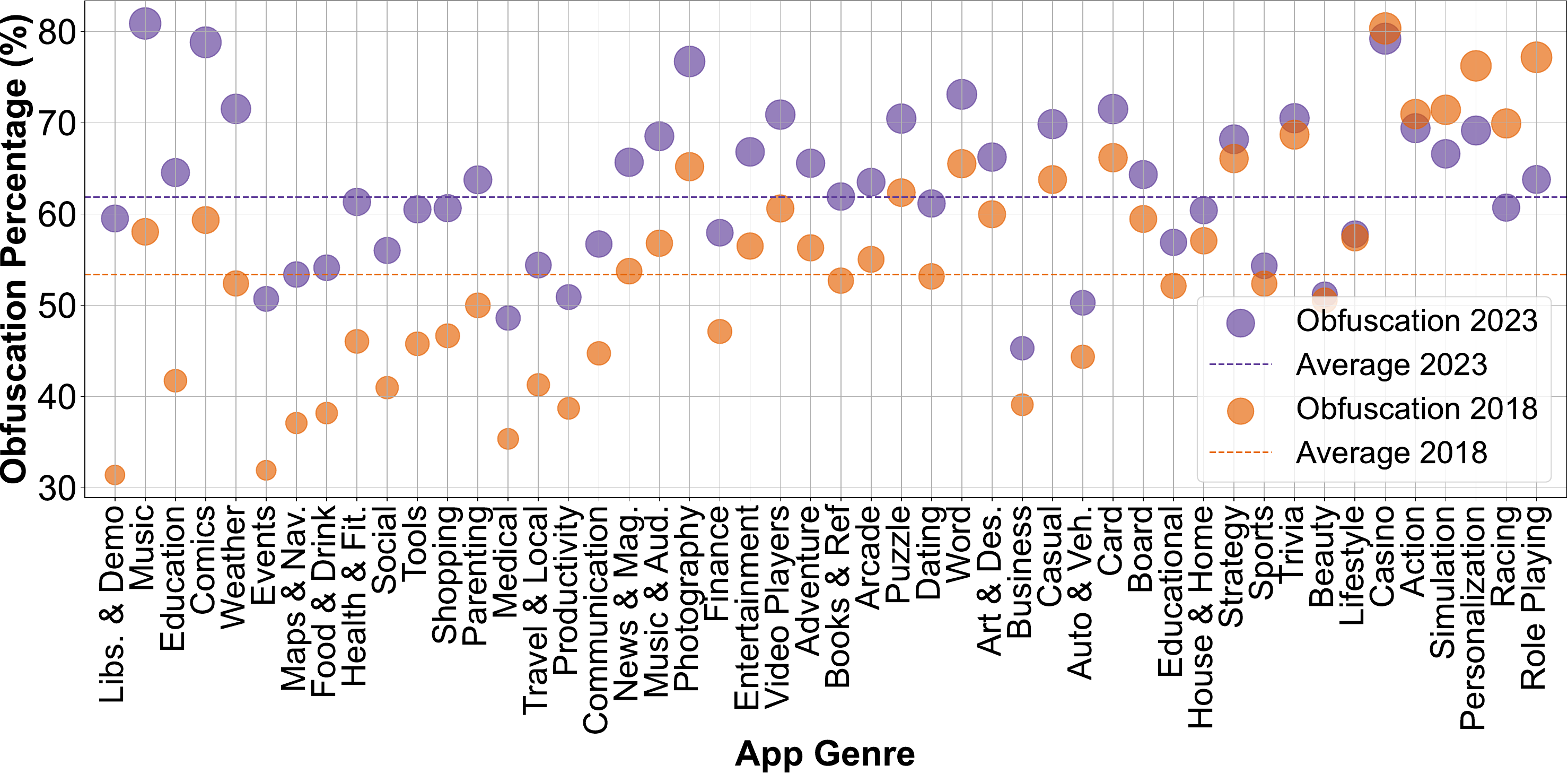}
    \caption{Percentage of obfuscated apps by genre (2018-2023)}
    \label{fig:app_genre_comparison}
\end{figure}

\subsubsection{Genre-wise obfuscation trends in the two snapshots} To investigate the adoption of obfuscation over time, we study the two snapshots of Google Play separately, i.e., APKs from 2016-2018 as one group and APKs from 2021-2023 as another. 

Figure~\ref{fig:app_genre_comparison} illustrates the change in obfuscation levels by app genre between 2016-2018 to 2021-2023. Notably, app categories such as Education, Weather, and Parenting, which had obfuscation levels below the 2018 average, have increased to above the 2023 average by 2023. One possible reason for this in Education and Parenting apps can be the increase in online education activities during and after COVID-19 and the developers identifying the need for app hardening.

There are some genres, such as Casino and Action, for which the percentage of obfuscated apps didn't change across the two snapshots (i.e., purple and orange circles are close together in Figure~\ref{fig:app_genre_comparison}). This is because these genres are highly obfuscated from the beginning. Finally, the four genres, including Simulation and Role Playing, have a lower percentage of obfuscated apps in the 2021-2023 dataset. Our manual analysis didn't result in a conclusion as to why.

\begin{figure}[!h]
    \centering
    \includegraphics[width=\linewidth]{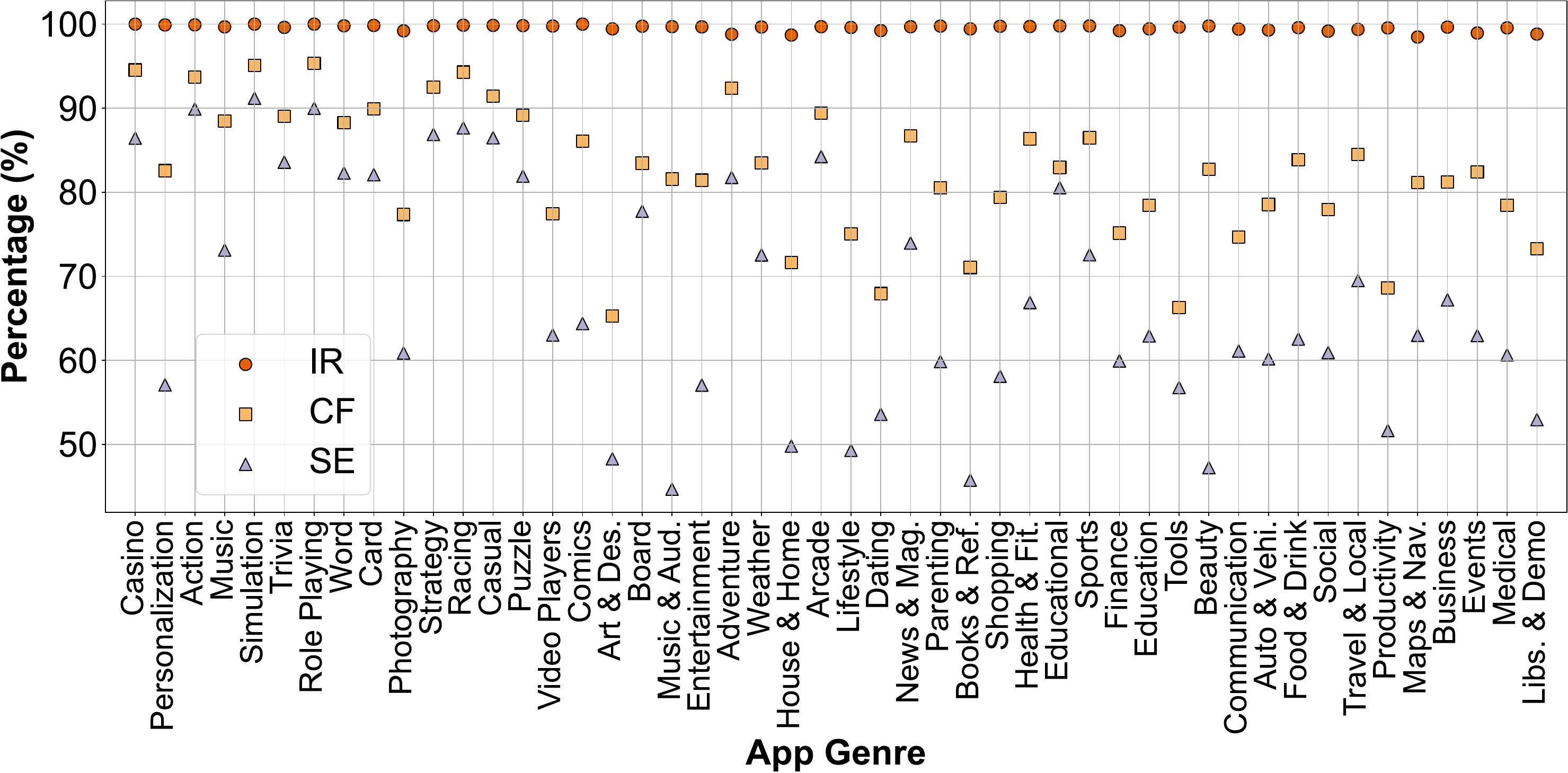}
    \caption{Obfuscation technique usage by genre (overall)}
    \label{fig:app_genre_all_tech}
\end{figure}

\subsubsection{Obfuscation techniques in different app genres} In Figure~\ref{fig:app_genre_all_tech}, we show the prevalence of key obfuscation techniques among various genres. As expected, almost all obfuscated apps in all genres used  Identifier Renaming. Also, it can be noted that genres with more obfuscated app percentages tend to use all three obfuscation techniques. Notably, more than 85\% of \textit{Casino} genre apps employ multiple obfuscation techniques

\subsubsection{Obfuscation tool usage in different app genres} We also investigated whether specific obfuscation tools are favoured by developers in different genres. However, apart from the expected observation that  ProGuard and Allatori being the most used tools, we didn't find any other interesting patterns. Therefore, we haven't included those measurement results.

\subsection{App Developers}
Next, we investigate individual developer-wise code obfuscation practices. From the pool of analyzed APKs, we identified the number of apps associated with each developer. Subsequently, we sorted the developers according to the number of apps they had created and selected the top 100 developers with the highest number of APKs for the 2016-2018 and 2021-2023 datasets. For the 2018 snapshot, we had 8,349 apps among the top 100 developers, while for the 2023 snapshot, we had 11,338 apps among the top 100 developers.

We then proceeded to detect whether or not these developers obfuscate their apps and, if so, what kind of tools and techniques they use. We present our results in five levels; developer obfuscating over 80\% of their apps, 60\%--80\% of apps, 40\%--60\% of apps, less than 40\%, and no obfuscation.

\begin{figure}[]
    \centering
    \includegraphics[width=\linewidth]{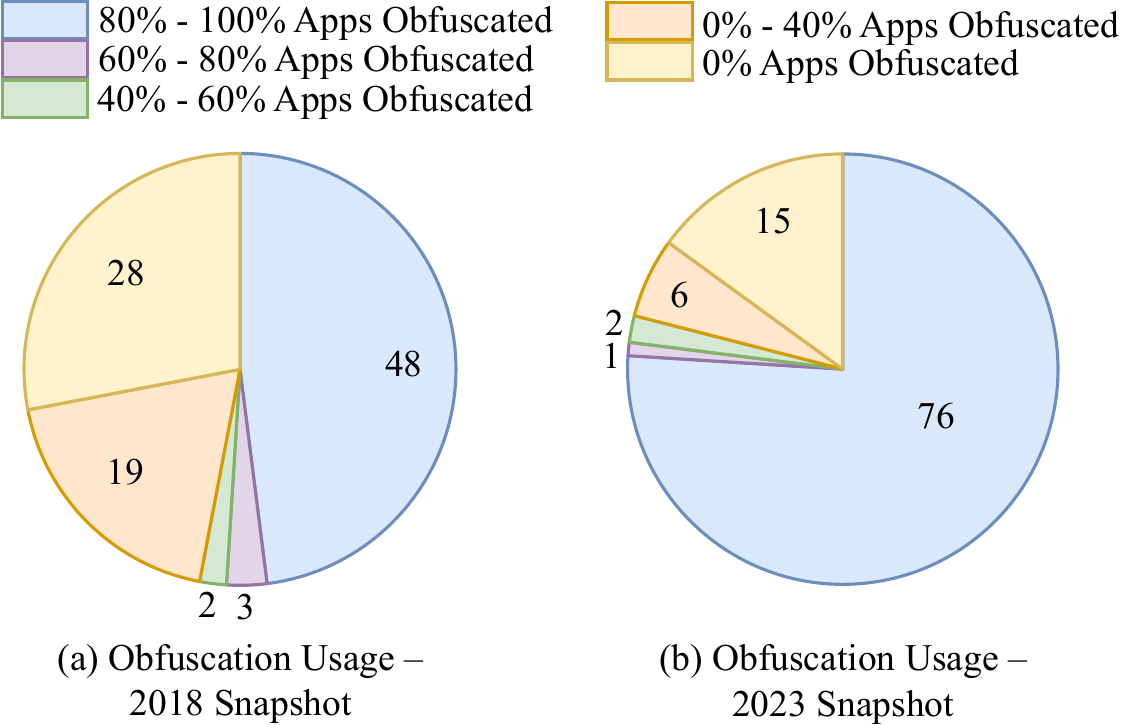}
    \caption{Obfuscation usage (Top-100 developers)}
    \label{fig:developer_trend_my_apps_all}
\end{figure}

Figure~\ref{fig:developer_trend_my_apps_all} compares the two datasets in terms of developer obfuscation adoption. It shows that more developers have moved to obfuscate more than 80\% of their apps in the 2021-2023 dataset (76\%) compared to the 2016-2018 dataset (48\%).

We also found that among developers who obfuscate more than 80\% of their apps, 73\% in 2018 and 93\% in 2023 used the same obfuscation tool. Additionally, these top developers employ Control Flow Modification (CF) and String Encryption (SE) above the average values discussed in Section~\ref{sec:obstrend}. Specifically, in 2018, top developers used CF in 81.3\% of cases and SE in 66.7\%, while in 2023, these figures increased to 88.2\% and 78.9\%. This results in two insights: 1) Most top developers obfuscate all their apps with advanced techniques, possibly due to concerns about IP and security, and 2) Developers stick to a single tool, possibly due to specialized knowledge or because they bought a commercial licence.

Finally, we investigate the obfuscation practices of developers with only one app in Table~\ref{tab:my-table}. According to the table, from those developers, 45.5\% of them obfuscated their apps in the 2016-2018 dataset and 57.2\% obfuscated their apps in the 2021-2023 dataset, showing a clear increase. However, these percentages are approximately 10\% lower than the average obfuscation rate in both cohorts discussed in Section~\ref{sec:obstrend}. This indicates that single-app developers may be less aware or concerned about code protection.

\begin{table}[]
\caption{Developers with only one app}
\label{tab:my-table}
\resizebox{\columnwidth}{!}{%
\begin{tabular}{cccccc}
\hline
\textbf{Year} & \textbf{\begin{tabular}[c]{@{}c@{}}Non\\ Obfuscated\end{tabular}} & \multicolumn{4}{c}{\textbf{Obfuscated}} \\ \hline
\multirow{3}{*}{\textbf{\begin{tabular}[c]{@{}c@{}}2018 \\ Snapshot\end{tabular}}} & \multirow{3}{*}{\begin{tabular}[c]{@{}c@{}}26,581 \\ (54.5\%)\end{tabular}} & \multicolumn{4}{c}{\begin{tabular}[c]{@{}c@{}}22,214 (45.5\%)\end{tabular}} \\ \cline{3-6} 
 &  & \textbf{ProGuard} & \textbf{Allatori} & \textbf{DashO} & \textbf{Other} \\ \cline{3-6} 
 &  & 6,131 & 8,050 & 658 & 7,375 \\ \hline
\multirow{3}{*}{\textbf{\begin{tabular}[c]{@{}c@{}}2023 \\ Snapshot\end{tabular}}} & \multirow{3}{*}{\begin{tabular}[c]{@{}c@{}}19,510 \\ (42.8\%)\end{tabular}} & \multicolumn{4}{c}{\begin{tabular}[c]{@{}c@{}}26,084 (57.2\%)\end{tabular}} \\ \cline{3-6} 
 &  & \textbf{ProGuard} & \textbf{Allatori} & \textbf{DashO} & \textbf{Other} \\ \cline{3-6} 
 &  & 12,697 & 9,672 & 234 & 3,581 \\ \hline
\end{tabular}%
}
\end{table}

\subsection{Top-k Apps}

Next, we investigate the obfuscation practices of top apps in Google Play Store. First, we rank the apps using the same criterion used by our previous work~\cite{rajasegaran2019multi, karunanayake2020multi, seneviratne2015early}. That is, we sort the apps in descending order of number of downloads, average rating, and rating count, with the intuition that top apps have high download numbers and high ratings, even when reviewed by a large number of users. Then, we investigated the percentage of obfuscated apps and obfuscation tools and technique usage as summarized in Table~\ref{tab:top_k_apps_2018_2023}.

When considering the highly ranked applications (i.e., top-1,000), the obfuscation percentage is notably higher, at around 93\%, in both datasets, which is significantly higher than the average percentage of obfuscation we observed in Section~\ref{sec:obstrend}. Top-ranked apps, likely due to their higher visibility and potential revenue, invest more in obfuscation to safeguard their intellectual property and enhance security. 

The obfuscation percentage decreases when going from the top 1,000 apps to the top 30,000 apps. Nonetheless, the obfuscation percentage in both datasets remains around similar values until the top 30,000 (e.g., $\sim$74\% for top-30,000). This indicates that the major increase in obfuscation in the 2021-2023 dataset comes from apps beyond the top 30,000.

When observing the tools used, the usage of ProGuard increases as we move from top to lower-ranked apps in both datasets. This may be because ProGuard is free and the default in Android Studio, while commercial tools like Allatori and DashO are expensive. There is a notable increase in the use of Allatori among the top apps in the 2021-2023 dataset. Regarding obfuscation techniques, the top 1,000 apps utilize all three techniques more frequently than lower-ranked apps in both snapshots. This indicates that the top 1,000 apps are more heavily protected compared to lower-ranked ones.

\begin{table*}[]
\caption{Summary of analysis results for Top-k apps in 2018 and 2023}
\label{tab:top_k_apps_2018_2023}
\resizebox{\textwidth}{!}{%
\begin{tabular}{lccccccccc}
\hline
\multicolumn{1}{c}{\begin{tabular}[c]{@{}c@{}}Top k apps - \\ Year\end{tabular}} & \begin{tabular}[c]{@{}c@{}}Total \\ Apps\end{tabular} & \begin{tabular}[c]{@{}c@{}}Obfuscation\\ Percentage\end{tabular} & \begin{tabular}[c]{@{}c@{}}ProGuard\\ Percentage\end{tabular} & \begin{tabular}[c]{@{}c@{}}Allatori\\ Percentage\end{tabular} & \begin{tabular}[c]{@{}c@{}}DashO\\ Percentage\end{tabular} & \begin{tabular}[c]{@{}c@{}}Other\\ Percentage\end{tabular} & \begin{tabular}[c]{@{}c@{}}IR\\ Percentage\end{tabular} & \begin{tabular}[c]{@{}c@{}}CF\\ Percentage\end{tabular} & \begin{tabular}[c]{@{}c@{}}SE\\ Percentage\end{tabular} \\ \hline
1k (2018) & 1,000 & 93.40 & 29.98 & 28.48 & 0.64 & 40.90 & 99.90 & 88.76 & 65.42 \\
10k (2018) & 10,000 & 85.19 & 25.55 & 35.32 & 0.47 & 38.65 & 99.90 & 88.76 & 71.91 \\
20k (2018) & 20,000 & 78.42 & 26.31 & 36.76 & 0.57 & 36.36 & 99.87 & 87.37 & 71.49 \\
30k (2018) & 30,000 & 74.40 & 27.30 & 37.71 & 0.64 & 34.36 & 99.82 & 86.75 & 71.11 \\
30k+ (2018) & 314,568 & 53.36 & 36.72 & 34.70 & 1.33 & 27.24 & 99.34 & 83.54 & 63.11 \\ \hline
1k (2023) & 1,000 & 92.50 & 24.00 & 51.89 & 1.95 & 22.16 & 100.0 & 92.54 & 83.68 \\
10k (2023) & 10,000 & 81.88 & 26.03 & 56.20 & 1.03 & 16.74 & 99.89 & 89.40 & 82.01 \\
20k (2023) & 20,000 & 76.62 & 30.48 & 52.92 & 0.96 & 15.64 & 99.93 & 85.80 & 78.01 \\
30k (2023) & 30,000 & 73.72 & 33.87 & 50.34 & 0.89 & 14.90 & 99.95 & 83.31 & 75.34 \\
30k+ (2023) & 206,216 & 61.90 & 46.56 & 38.21 & 0.64 & 14.59 & 99.97 & 77.51 & 62.50 \\ \hline
\end{tabular}%
}
\end{table*}

\section{Related Work}
\label{Sec:Related Work}

\subsection{Obfuscation Detection}

Multiple works developed methods to detect obfuscation and identify the tools and techniques used. For instance, Kühnel et al.~\cite{kuhnel2015fast} introduced the IREA framework, employing a rule-based detection algorithm tailored to identify obfuscation techniques such as Identifier Renaming. Despite achieving high accuracy, this approach's reliance on specific rules limits its applicability across diverse scenarios. 

Similarly, Wermke et al.~\cite{wermke2018large} developed OBFUSCAN to emulate ProGuard's behaviour, focusing primarily on detection techniques aligned with ProGuard's functionalities. However, this specialization restricts its performance with other obfuscation tools. In contrast, Wang et al.~\cite{wang2017changed} presented a multi-class classifier utilizing SVM to identify obfuscation tools and configurations, demonstrating promising accuracy levels. Nonetheless, the limited dataset used for validation raises concerns regarding its generalizability. Dong et al.~\cite{dong2018understanding} and Park et al.~\cite{park2019framework} proposed machine learning-based methods to detect obfuscation techniques. AndrODet~\cite{mirzaei2019androdet} and AndrODet*~\cite{conti2022obfuscation} also had similar approaches.

In our work, we draw ideas from these machine learning-based obfuscation detectors and design a comprehensive framework that facilitates large-scale analysis of APKs. Also, in contrast to these works, we train and test our classifiers in a diverse set of datasets to increase the generalizability that is required for real-world settings.

\subsection{Empirical Studies of Obfuscation}
OBFUSCAN~\cite{wermke2018large} is the most related to our work which examined obfuscation usage in the Google Play Store and developer awareness through a survey with 1.7 million apps from 2010 to 2017, primarily detecting Identifier Renaming. 
Similarly, Dong et al.~\cite{dong2018understanding} introduced an obfuscation detection methodology and conducted a large-scale investigation using 26k Google Play Store apps and 65k 3rd Party Apps from 2016 to 2017. However, the authors mainly focus their study on obfuscation techniques.

Another study~\cite{wang2018software} introduced an NLP-based obfuscation detection tool focusing on symbol renaming and conducted a large-scale empirical study on the iPhone Operating System. Hammad et al.~\cite{hammad2018large} discussed the effects of code obfuscations in Android apps, evaluating commercial anti-malware products against various obfuscation tools and strategies. Additionally, Kargén et al.~\cite{kargen2023characterizing} used anomaly detection followed by manual inspection to investigate popular obfuscation techniques among Malware and Google Play Store APKs, mainly focusing on Control Flow obfuscation with Java Reflections using 13k apps released in 2020.

\textit{In contrast to these works, our research is the largest of its kind, covers a span of eight years, allowing us to observe trends, and covers a broader scope of code obfuscation practices simultaneously}.

\section{Discussion and Concluding Remarks}
\label{sec:discussion}

Using our obfuscation detection framework, we conducted a large-scale, eight-year investigation into code obfuscation practices in the Google Play Store, analyzing more than 500,000 APKs. To the best of our knowledge, this study is the first of its kind. Finally, we discuss the implications and limitations of our findings.

\subsection{Implications}

\noindent{{\bf Adoption of code obfuscation:}} Overall, code obfuscation is on an increasing trend in the Google Play Store. More specifically, we found that the average percentage of obfuscated apps between 2016-2018 was 53\%, and that increased to 62\% in 2021-2023. These results indicate more and more developers are aware of the associated intellectual property and security issues in Google Play and are taking actions to mitigate them. However, app store administrators may want to balance obfuscation and readability since excessive code obfuscation can hinder in-build security checks in app stores, e.g., Bouncer~\cite{nawaz2022evaluation, bacci2018impact}. As a result, understanding existing obfuscation practices will allow app store admins to build policies that balance obfuscation and code understandability. \\ \vspace{-3mm}

\noindent{{\bf Use of code obfuscation tools:}} We found that Proguard is the most commonly used obfuscation tool (40.92\%), likely because it is free and the default option in Android Studio. Surprisingly, a significant fraction of apps use the commercial obfuscator Allatori (36.64\%). Additionally, we found that 21.43\% of apps use unknown obfuscation tools, which presents a potential direction for future research. This trend may be driven by developers seeking advanced protection for their intellectual property, opting for more advanced tools like DexGuard~\cite{dexguard} over more commonly used options such as ProGuard. In addition, these empirical findings are important because often malware analysts conduct code reverse engineering and must be aware of available obfuscation tools and techniques~\cite{zhang2021android, wang2017changed, kuhnel2015fast}. Our results, including details of commonly used tools and the possible existence of non-mainstream/unknown obfuscators, will interest them. \\ \vspace{-3mm}

\noindent{{\bf Obfuscation techniques:}} Our results showed that Identifier Renaming is the most common obfuscation technique, used by 99.62\% of apps. We also found that 58.7\% of apps use all three main obfuscation techniques. The use of multiple obfuscation techniques, rather than a single technique, introduces additional complexity to the obfuscated app, thereby making reverse engineering more challenging. This highlights the need for more complex de-obfuscation solutions for app analysis as a necessary future research direction. Current de-obfuscation methods, such as those proposed in~\cite{bichsel2016statistical, baumann2017anti, you2022deoptfuscator}, tend to focus on individual obfuscation techniques, which limits their scope. Given the evolving industry trends, it is essential to address more complex combinations of obfuscation techniques in future. \\ \vspace{-3mm}

\noindent{{\bf App genres:}} Our app genre-wise obfuscation analysis found that Gaming and Casino apps use obfuscation more frequently than others. This is indeed not surprising. Due to financial transactions and gambling, Casino apps strive for obfuscation. The competitive nature and vulnerability to re-packing in gaming apps also explain their higher obfuscation usage. In future, we can expect other app categories that also handle important data and transactions, such as Finance, Health and Fitness, and Medical, to adopt more obfuscation. However, it is important to note that the developers must not use obfuscation as a security solution, as many examples in the past have shown that ``security by obscurity'' doesn't work. \\ \vspace{-3mm}

\noindent{{\bf Top developers and top apps:}} Finally, we found that top developers and apps use obfuscation more frequently, often preferring commercial obfuscators for added protection. This highlights the importance of obfuscation tools for app developers, encouraging their use to protect apps and IP~\cite{faruki2016android}. Furthermore, it educates small-scale developers on best practices from top developers and can help establish industry standards for obfuscation, especially in the era of GenAI, where code data are used to train AI models, sometimes without developer consent. To a certain extent, longitudinal data demonstrate an increasing adoption of obfuscation by smaller developers, which is a positive indicator of the overall health of the Android ecosystem. \\ \vspace{-7mm}

\subsection{Limitations and Future Work}
\label{sec:limitations}
\noindent{{\bf Limitations:}} In our large-scale analysis, we leveraged two separate datasets collected in 2018 and 2023. As a result, we didn't have sufficient representative samples from 2019 and 2020. While this is a limitation of our analysis, overall trends we observed are unlikely to change even including those data.

It is also important to note that the classifiers discussed in Section~\ref{sec:classifier_bank} could have prediction errors, which may propagate into our large-scale analysis, potentially influencing the overall findings. To mitigate this, we validated our classifiers' performance on unseen data ({\bf cf.} Section~\ref{sec:training_and_validation}), demonstrating their generalizability. However, there can still be error propagation, and with real-world APKs, there is no method for accurate ground-truth establishment. 
\\

\noindent{{\bf Future Work:}} In our obfuscation detection process, we classify APKs as a whole. However, obfuscation may originate from specific libraries within the APK, whether first-party or third-party. A potential future research direction is to examine which parts of the app are obfuscated. Conducting such an analysis on real-world data without ground truth is challenging, as it becomes difficult to identify the exact library once it has been obfuscated. This challenge is commonly encountered in third-party library detection research in Android~\cite{wang2018orlis, zhan2020automated, zhang2019libid, wu2023libscan}.

Additionally, our analysis focused on three commonly used and accessible tools. Some apps may employ DexGuard ({\bf cf.} Section~\ref{sec:commontools}), and our classifier should ideally categorize such apps as ‘Others.’ However, due to similarities between DexGuard and ProGuard, apps using DexGuard might be mistakenly classified as using ProGuard. A possible future direction is to extend this research by incorporating additional obfuscators, including commercially licensed tools like DexGuard, to improve the classifier’s capability and address this limitation.

\section*{Acknowledgments}
This research was supported by the Australian Government through the Australian Research Council’s Discovery
Projects funding scheme (Project ID DP220102520).

\bibliographystyle{IEEEtran}
\bibliography{references}

\newpage
\section{Biographies}
\begin{IEEEbiography}[{\includegraphics[width=1in,height=1.25in,clip,keepaspectratio]{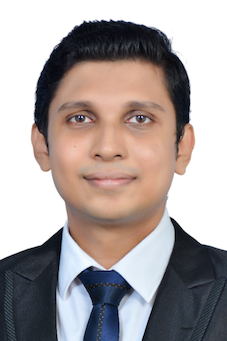}}]{Akila Niroshan} (Student  Member, IEEE) is currently pursuing a PhD at the School of Electrical Engineering and Telecommunications, University of New South Wales. His research interests include Android code obfuscation and AI-enhanced solutions for mobile app security. Before moving into research, he worked for nearly five years in the telecommunications industry, focusing on the Internet of Things. He received his B.Sc. (Hons) in Electronics and Telecommunications Engineering from the University of Moratuwa, Sri Lanka, in 2017.
\end{IEEEbiography}

\begin{IEEEbiography}[{\includegraphics[width=1in,height=1.25in,clip,keepaspectratio]{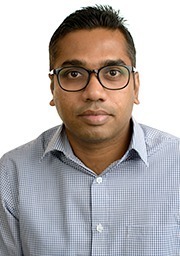}}]{Suranga Seneviratne} (Senior Member, IEEE) is a Lecturer in Security at the School of Computer Science, The University of Sydney. He received his PhD from the University of New South Wales, Australia in 2015. His current research interests include privacy and security in mobile systems, AI applications in security, and behavior biometrics. Before moving into research, he worked for nearly six years in the telecommunications industry in core network planning and operations. He received his bachelor's degree from the University of Moratuwa, Sri Lanka in 2005.
\end{IEEEbiography}

\begin{IEEEbiography}[{\includegraphics[width=1in,height=1.25in,clip,keepaspectratio]{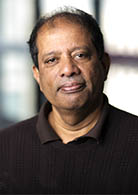}}]{Aruna Seneviratne} (Senior Member, IEEE) is currently a Foundation Professor of Telecommunications with the University of New South Wales, Australia, where he holds the Mahanakorn Chair of Telecommunications. He has also worked at a number of other universities in Australia, U.K., and France, and industrial organizations, including Muirhead, Standard Telecommunication Labs, Avaya Labs, and Telecom Australia (Telstra). He has held visiting appointments with INRIA, France. His current research interests are in physical analytics: technologies that enable applications to interact intelligently and securely with their environment in real-time. Most recently, his team has been working on using these technologies in behavioral biometrics, optimizing the performance of wearables, and IoT system verification. He has been awarded a number of fellowships, including one at British Telecom and one at Telecom Australia Research Labs.
\end{IEEEbiography}

\vfill

\end{document}